\documentclass{iaus}
\usepackage{graphicx}

\newcommand{\beq}{\begin{equation}}
\newcommand{\eeq}{\end{equation}}
\def\farcm{\hbox{$.\mkern-4mu^\prime$}}

\def\solar{\mbox{$_{\normalsize\odot}$}}

\newcommand{\lsim}{\ \raise
-2.truept\hbox{\rlap{\hbox{$\sim$}}\raise5.truept\hbox{$<$}\ }}
\newcommand{\gsim}{\ \raise
-2.truept\hbox{\rlap{\hbox{$\sim$}}\raise5.truept\hbox{$>$}\ }}
\newcommand{\simsim}{\ \raise
-2.truept\hbox{\rlap{\hbox{$\sim$}}\raise5.truept\hbox{$\sim$}\ }}

\title[The sub-solar IMF in the Large Magellanic Cloud] {The sub-solar
Initial Mass Function\\ in the Large Magellanic Cloud\thanks{Based on
observations made with the NASA/ESA {\em Hubble Space Telescope},
obtained at the Space Telescope Science Institute, which is operated by
the Association of Universities for Research in Astronomy, Inc. under
NASA contract NAS 5-26555.}}

\author[Dimitrios A. Gouliermis]{Dimitrios A. Gouliermis}

\affiliation{Max-Planck-Institut f\"{u}r Astronomie, \\ K\"{o}nigstuhl
17, 69117 Heidelberg, Germany \\ email: {\tt dgoulier@mpia.de}}

\pubyear{2008}
\volume{256}  
\pagerange{119--126}
\jname{The Magellanic System: Stars, Gas, and Galaxies}
\editors{Jacco Th. van Loon \& Joana M. Oliveira, eds.}
\begin{document}

\maketitle

\begin{abstract}

The Magellanic Clouds offer a unique variety of star forming regions
seen as bright nebulae of ionized gas, related to bright young stellar
associations. Nowadays, observations with the high resolving efficiency
of the Hubble Space Telescope allow the detection of the faintest infant
stars, and a more complete picture of clustered star formation in our
dwarf neighbors has emerged. I present results from our studies of the
Magellanic Clouds, with emphasis in the young low-mass pre-main sequence
populations. Our data include imaging with the Advanced Camera for
Surveys of the association LH~95 in the Large Magellanic Cloud, the
deepest observations ever taken with HST of this galaxy. I discuss our
findings in terms of the Initial Mass Function, which we constructed
with an unprecedented completeness down to the sub-solar regime, as the
outcome of star formation in the low-metallicity environment of the LMC.


\end{abstract}

\firstsection % if your document starts with a section,
\section{Introduction}

The conversion of gas to stars is determined by the star formation
process, the outcome of which are stars with a variety of masses. The
distribution of stellar masses in a given volume of space at the time of
their formation is known as the Initial Mass Function (IMF), given that
all stars were born simultaneously. The IMF dictates the evolution and
fate of stellar systems, as well as of whole galaxies. The evolution of
a stellar system is driven by the relative initial numbers of stars of
various masses, from the short-lived high-mass stars ($M$ \gsim\ 8
M{\solar}), which enrich the ISM with elements heavier than H and He, to
the low-mass stars ($M$ \lsim\ 1 M{\solar}), which lock large amounts of
mass over long timescales. It is therefore of much importance to
quantify the relative numbers of stars in different mass ranges and to
identify systematic variations of the IMF with different star-forming
conditions, which will allow us to understand the physics involved in
assembling each of the mass ranges.

There are various parameterizations of the IMF (see \cite{kroupa02} and
\cite{chabrier03} for reviews), of which a commonly used is that of a
power law of the form $\xi(\log{M}) \propto M^{\Gamma}$, or
alternatively $\xi(M) \propto M^{\alpha}$.  The IMF, thus, is
characterized by the derivatives $$ \Gamma =
\frac{d\log{\xi(\log{M})}}{d\log{M}}~~~{\rm or}~~~\alpha =
\frac{d\log{\xi(M)}}{d\log{M}}{\rm ,}\label{eq:IMF} $$ depending on
whether stars are distributed according to their masses in logarithmic
or linear scales. The above derivatives, which relate to each other as
$\alpha = \Gamma -1$, correspond to the so-called slope of the IMF. A
reference value for the IMF slope, as found by \cite{salpeter55} in the
solar neighborhood for stars with masses between 0.4 and 10 M{\solar},
is $\Gamma=-1.35$ (or $\alpha = -2.35$). In general, although the IMF
appears relatively uniform when averaged over whole clusters or large
regions of galaxies (\cite{chabrier03}), the measured IMF shows local
spatial variations, which could be the result of physical differences,
or even purely statistical in nature (\cite{elmegreen04}). 

\begin{figure}[t!]
% \vspace*{-2.0 cm}
\begin{center}
\includegraphics[width=0.975\textwidth]{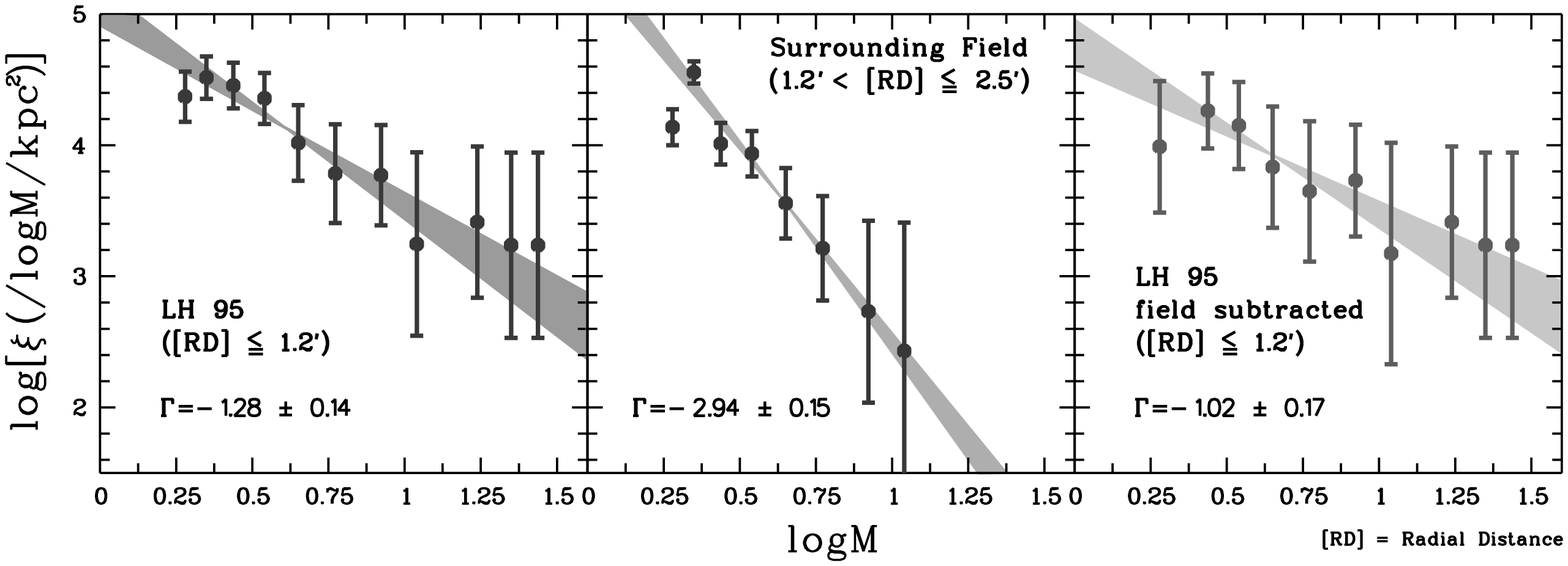} 
% \vspace*{-1.0 cm}
\caption{The main-sequence MF of the association LH~95 from observations
with the 1-m telescope at Siding Spring Observatory. {\em Left}: The MF
of the system (radial distance $r$~\lsim~1\farcm2 from its center) with
the field population included. {\em Middle}: The main-sequence MF of the
considered surrounding field within radial distance
1\farcm2~\lsim~$r$~\lsim~2\farcm5. {\em Right}: The main-sequence MF of
the association after the field contribution has been subtracted, which
accounts for the IMF of the system. The lack of stars with masses
\gsim~10~M{\solar} in the surrounding field introduces a steep MF for
its stellar population. The shaded regions represent the uncertainties
of the MF. The corresponding MF slopes $\Gamma$ are given for stars up
to the higher observed mass, which in the region of the system is
$\sim$~28~M{\solar}. The slope of the IMF of LH~95 is a bit shallower
than a \cite{salpeter55} IMF because of its high-mass content. Data
from \cite{gouliermis02}. \label{f:lh95imf}}
\end{center}
\end{figure}

The Large Magellanic Cloud (LMC) is the nearest undisrupted neighboring
dwarf galaxy to our own. It has a spatially varying sub-solar
metallicity of $Z \sim$~0.3~-~0.5~$Z${\solar}, while its star formation
rate is almost the same with the Milky Way (\cite[Westerlund
1997]{westerlund97}). It demonstrates an energetic star formation
activity with its {\sc Hi} shells (\cite[Kim et al. 1999]{kim99}), {\sc
Hii} regions (\cite[Davies et al. 1976]{davies76}), and molecular clouds
(\cite[Fukui et al. 1999]{fukui99}), all linked to ongoing star
formation. A wide variety of young stellar systems, the stellar
associations (\cite[Gouliermis et al. 2003]{gouliermis03}), located at
regions of recent star formation in the LMC form a complete sample of
targets with various characteristics for the study of the stellar IMF in
this galaxy. Therefore, the LMC, being so close to us, has provided an
ideal alternative environment for the study of extragalactic star
formation and the derived IMF. 

Photometric and spectroscopic investigations of young stellar
associations in the LMC, limited so far to ground-based observations,
revealed that the IMF of the high-mass stars in these systems is
consistent from one system to the other. All measured IMF slopes are
found to be clustered around the Salpeter value, not very different from
that of OB associations of our Galaxy (\cite[Massey 2006]{massey06}),
suggesting that the massive IMF appears more or less to be universal
with a typical Salpeter slope (\cite[Massey 2003]{massey03}). It should
be noted that there are small but observable differences in the
constructed IMF and its slope between different systems, which may imply
that the IMF is probably determined by the local physical conditions
(\cite[Hill et al. 1995]{hill95}), or that the variability in the slope
of the IMF could be the result of different star formation processes
(\cite[Parker et al. 1998]{parker98}). However, taking into account the
constraints in the construction of the IMF, one may argue that the IMF
variations are possibly observational and/or statistical in nature
(\cite[Kroupa 2001]{kroupa01}). An example of the spatial dependence of
the IMF and the effect of the contamination by the field population is
demonstrated in Fig.~\ref{f:lh95imf}, where the massive IMF in the LMC
association LH~95 constructed earlier by the author and collaborators is
shown.

\section{The IMF in the Low-Mass Regime}

Complete investigations concerning the low-mass stars are so far
available only for the Galactic IMF, which is found to become flat in
the sub-solar regime (\cite[Reid 1998]{reid98}), down to the detection
limit of 0.1~M{\solar} or lower (\cite{lada98}). Observed variations
from one region of the Galaxy to the next in the numbers of low-mass
stars and brown dwarfs over intermediate- and high-mass stars affect the
corresponding IMF, which seems to depend on the position (e.g.
\cite{luhman00}; \cite{briceno02}; \cite{preibisch03}). Explanations
suggested for the low-mass IMF variations include stochasticity in the
ages and ejection rates of proto-stars from dense clusters (e.g.
\cite{bate02}), differences in the photo-evaporation rate from high-mass
neighboring stars (e.g. \cite{kroupa03}), or a dependence of the Jeans
mass on column density (\cite{briceno02}) or Mach number
(\cite{padoan02}), and they may also be affected by variations in the
binary fraction (\cite{malkov01}). 

While the discussion over the low-mass part of the Galactic IMF
continues, more information on the IMF toward smaller masses in other
galaxies is required for the understanding of its dependence on the
global properties of the host-galaxy and for addressing the issue of its
universality. The LMC, due to its proximity, can serve as the best proxy
for the study of extragalactic young low-mass stellar populations, and
due to its global differences from the Milky Way, it provides a
convenient alternative environment for the investigation of the IMF
variability. Under these circumstances, issues that should be addressed
can be summarized to {\em what is the low-mass stellar population of
young stellar associations in the LMC}, {\em what is the shape of the
corresponding IMF}, and {\em how it compares to that of the Milky Way}.
A substantial amount of such information in both the Magellanic Clouds
(MCs) was thus far lacking, mostly due to observational limitations.
However, recent observations of the MCs with the {\em Hubble Space
Telescope} ({\em HST}) changed dramatically our view of star formation
in these galaxies.

\begin{figure}[t!]
\begin{center}
\includegraphics[width=0.875\textwidth]{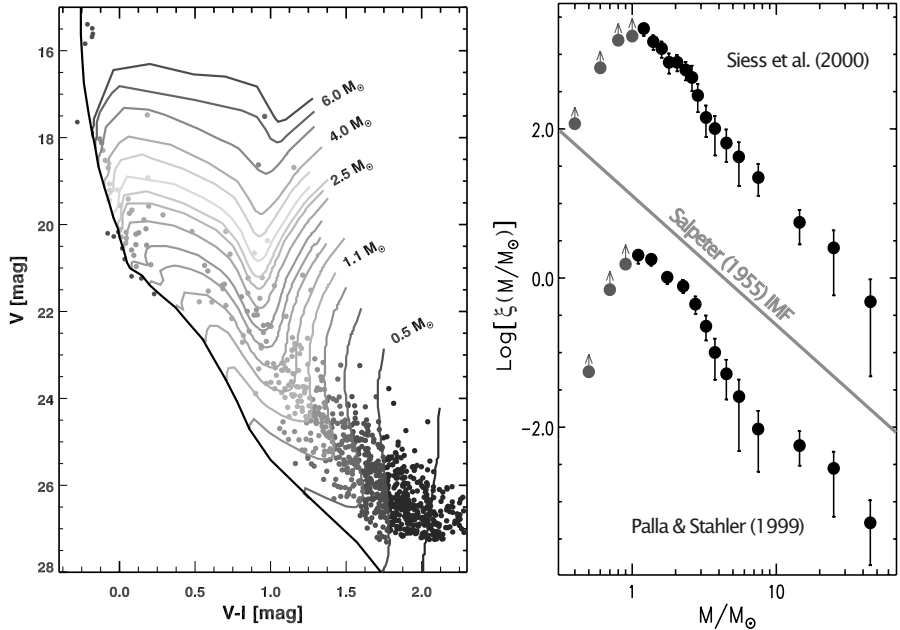} 
\caption{Construction of the IMF of the young SMC cluster NGC~602. {\em
Left}: The CMD of the stars of the system after the contamination by the
field population is removed, with PMS evolutionary tracks from the
models by \cite{siess00} overlaid. The ZAMS from the models by
\cite{girardi02} is also shown.  The IMF of NGC~602 for the PMS stars is
constructed by counting stars between evolutionary tracks using both
\cite{palla99} and \cite{siess00} PMS tracks, while the ZAMS is used for
the construction of a mass-luminosity relation for the bright MS stars
with $M \gsim 6$~M{\solar} for the derivation of their masses. {\em
Right}: The derived IMF, as it is constructed with use of both the
models by \cite{siess00} (top) and \cite{palla99} (bottom). The grey
line corresponds to a \cite{salpeter55} IMF with $\alpha\simeq-2.3$.
Stellar numbers are corrected to a bin-size of 1~M{\solar}. Data from
\cite{schmalzl08}.\label{f:pms-imf}}
\end{center}
\end{figure}

\section{Low-Mass PMS Populations in the Magellanic Clouds}

In the Galaxy previous observations have confirmed the existence of
faint pre-main sequence (PMS) stars as the low-mass population of nearby
OB associations and star-forming regions (e.g. \cite{hillenbrand93};
\cite{brandl99}; \cite{preibisch02}; \cite{sherry04}). The only
extragalactic places where such stars could be resolved are the MCs, and
recent imaging with {\em HST} revealed the PMS populations of their
stellar associations, and allowed their study. Our investigation of the
LMC association LH~52 with WFPC2 observations extended the stellar
membership of MCs associations to their PMS populations for the first
time (\cite{gouliermis06a}). Subsequent photometry with the {\em
Advanced Camera for Surveys} (ACS) of the association NGC~346 in the
Small Magellanic Cloud (SMC) led to the discovery of an extraordinary
number of low-mass PMS stars in its vicinity (\cite{nota06};
\cite{gouliermis06b}), providing the required statistical sample for the
investigation of the clustering behavior (\cite{hennekemper08}) and the
IMF (\cite{sabbi08}) of these stars.

Our photometric study of another young SMC star cluster, NGC~602, with
{\em HST}/ACS revealed a coherent sample of PMS stars, ideal for the
study of the low-mass IMF in the SMC and the complications in its
construction (\cite[Schmalzl et al. 2008]{schmalzl08}). The IMF of the
PMS population of NGC~602 was constructed by counting the stars between
evolutionary tracks on the CMD (Fig.~\ref{f:pms-imf} - {\em left}).
While this method is independent of any age-gradient among the stars, it
still depends on the selected models. As a consequence, the shape of the
derived IMF for stars with $M$~\lsim~2~M{\solar} appears somewhat
different with the use of different PMS tracks (Fig.~\ref{f:pms-imf} -
{\em right}). Although this IMF seems to flatten for masses close to
solar, no important change in the slope is identified. We found that, in
general, a single-power law with a slope of $\alpha\simeq-2.2\pm0.3$
represents well the average IMF of the system as it is constructed with
the use of all considered grids of models (\cite[Schmalzl et al.
2008]{schmalzl08}). This result is in line with that of \cite[Sabbi et
al. (2008)]{sabbi08} for NGC~346. The IMF of both NGC~346 and NGC~602 is
limited to stars of 1~M{\solar} due to the observations, which did not
allow the detection of statistically significant numbers of sub-solar
PMS stars. 

\begin{figure}[t!]
% \vspace*{-2.0 cm}
\begin{center}
\includegraphics[width=0.975\textwidth]{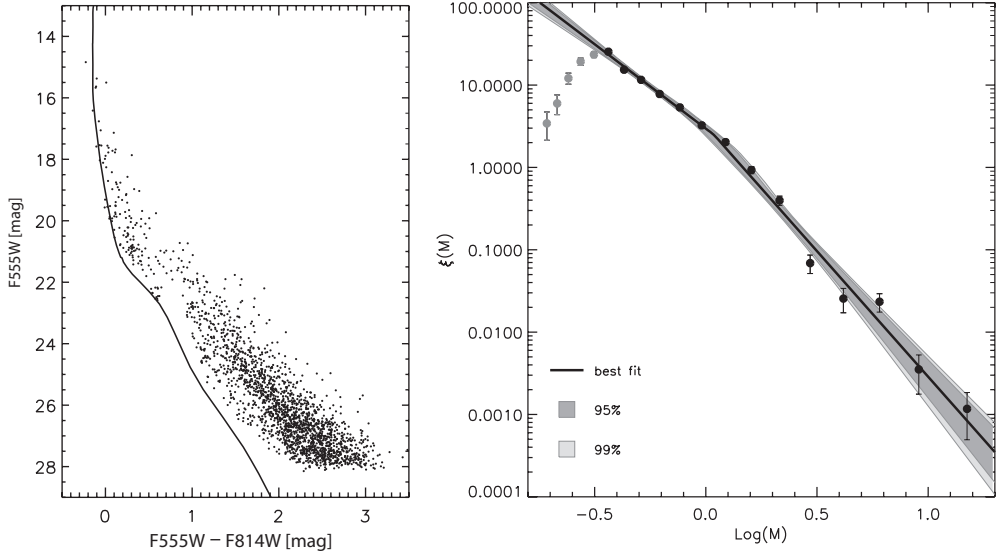} 
% \vspace*{-1.0 cm}
\caption{{\em Left}: The F555W$-$F814W, F555W CMD of the stars detected
with {\em HST}/ACS in the LMC association LH~95, after the contamination
of the LMC field has been statistically subtracted (\cite[Gouliermis et
al. 2007]{gouliermis07}). The ZAMS from \cite{girardi02} grid of models
is also plotted. This is the most complete CMD of an extragalactic
star-forming region ever constructed. {\em Right}: The IMF of the
association LH~95 constructed from our ACS photometry with the use of a
new observable plane for the PMS models by \cite{siess00} designed by us
for the metallicity of the LMC and the photometric system of ACS. The
best two-part power-law fit is drawn with solid line, while the shaded
areas represent the $95\%$ and $99\%$ confidence uncertainties in the
slope determination and the break point (the ``knee''). Units of the
IMF, $\xi(M)$, are logarithmic number of stars per solar mass per
pc$^2$. Data from \cite{dario09}.\label{f:ss-imfs}}
\end{center} 
\end{figure}

\section{The sub-solar IMF in the Large Magellanic Cloud}

The first complete sample of extragalactic sub-solar PMS stars is
detected by the author and collaborators by utilizing the {\em high
sensitivity and spatial resolving power} of ACS in combination with its
{\em large field of view} in the LMC association LH~95 (\cite[Gouliermis
et al. 2007]{gouliermis07}). Two pointings, one on LH~95 and another on
the general field, were observed in the filters F555W ($\approx V$) and
F814W ($\approx I$) with the longest exposures ever taken with {\em HST}
of the LMC (in total 5,000 sec per filter per field). These
state-of-the-art observations allowed us the construction of the CMD of
the association in unprecedented detail and to decontaminate it for the
average LMC stellar population (Fig.~\ref{f:ss-imfs} - {\em left}).
Although LH~95 represents a rather modest star-forming region, our
photometry, with a detection limit of $V$~\lsim~28~mag (at 50\%
completeness), revealed more than 2,500 PMS stars with masses down to
$\sim$~0.3~M{\solar}. 

The subsequent interpretation of these observations by \cite{dario09}
led to the introduction of a new set of observational evolutionary
models, derived from the theoretical calculations by \cite{siess00}. We
converted luminosities from these evolutionary tracks into observable
magnitudes in the ACS photometric system by taking into account the
parameters $T_{\rm eff}$, $\log{(g)}$ and $[M/H]$ for a large variety of
stellar objects. We used these models with the observations of LH~95 to
derive the IMF of the system by assigning a mass value to each PMS star.
This IMF, shown in Fig.~\ref{f:ss-imfs} ({\em right}), is reliably
constructed for stars with masses down to $\sim 0.3$~M$_{\odot}$, the
lowest mass ever observed within reasonable completeness in the MCs.
Consequently, its construction offers {\em an outstanding improvement in
our understanding of the low-mass star formation in the LMC}. We
verified statistically that the field-subtracted completeness-corrected
IMF of LH~95 has a definite change in its slope for masses
$M$~\lsim~1~M{\solar}, where it becomes more shallow (\cite[Da Rio,
Gouliermis \& Henning 2009]{dario09}). In general, the shape of this IMF
agrees very well with a multiple power-law, as the typical Galactic IMF
down to the sub-solar regime. A definite change is identified, though,
in the slope of the IMF (the ``knee of the IMF'') at $\sim 1$~M{\solar},
a higher mass-limit than that of the average Galactic IMF derived in
various previous investigations (e.g. \cite[Kroupa 2001]{kroupa01},
\cite[2002]{kroupa02}). As far as the slope of the IMF of LH~95 is
concerned, it is found to be systematically steeper than the classical
Galactic IMF (e.g. \cite[Scalo 1998]{scalo98}; \cite[Kroupa
2002]{kroupa02}) in both low- and high-mass regimes. No significant
differences in the shape of the average IMF of LH~95 from that of each
of the three individual PMS sub-clusters of the association is also
found. This clearly suggests that the IMF of LH~95 is not subject to
local variability. 

\begin{acknowledgments}
The author kindly acknowledges the support of the German Research
Foundation (DFG) through the individual grant GO 1659/1-1. 
\end{acknowledgments}

\end{document}